\documentstyle[12pt]{article}
\newcommand\bea{\begin{eqnarray}}
\newcommand\eea{\end{eqnarray}}
\begin{document}
\bibliographystyle{unsrt}
\setlength{\baselineskip}{18pt}
\parindent 18pt

\begin{center}{
{\Large {\bf Deformed quantum harmonic oscillator \\
with diffusion and dissipation
} } \vskip 1truecm
A. Isar${\dagger\ddagger}^{(a)}$
and W. Scheid${\ddagger}$\\ $\dagger${\it Department of Theoretical
Physics, Institute of Physics and Nuclear Engineering\\ Bucharest-Magurele,
Romania }\\
$\ddagger${\it Institut f\"ur Theoretische Physik der
Justus-Liebig-Universit\"at \\ Giessen, Germany }\\ }
\end{center}

\begin{abstract}
A master equation for the deformed quantum harmonic oscillator
interacting with a dissipative environment, in particular with a
thermal bath, is derived in the microscopic model by using
perturbation theory. The coefficients of the master equation and
of equations of motion for observables depend on the deformation
function. The steady state solution of the equation for the
density matrix in the number representation is obtained and the
equilibrium energy of the deformed harmonic oscillator is
calculated in the approximation of small deformation.
\end{abstract}


(a) e-mail address: isar@theory.nipne.ro

\section{Introduction}

For more than a decade a constant interest has been induced to the
study of deformations of Lie algebras -- so-called  quantum
algebras or quantum groups, whose rich structure produced
important results and consequences in statistical mechanics,
quantum field theory, conformal field theory, quantum and
nonlinear optics, nuclear and molecular physics. Their use in
physics became intense with the introduction in 1989, by
Biedenharn \cite{bied} and MacFarlane \cite{macf}, of the
$q$-deformed Heisenberg-Weyl algebra ($q$-deformed quantum
harmonic oscillator). Since then the properties of the
deformations of the harmonic oscillator have been investigated by
many authors. Several kinds of generalized deformed oscillators
have been introduced. There are, at least, two properties which
make $q$-oscillators interesting objects for physics. The first is
the fact that they naturally appear as the basic building blocks
of completely integrable theories. The second concerns the
connection between $q$-deformation and nonlinearity. In Refs.
\cite{mank2,mank1,ani} it was shown that the $q$-oscillator leads
to nonlinear vibrations with a special kind of the dependence of
the frequency on the amplitude. For example, the $q$-deformed Bose
distribution  produces a correction to the Planck distribution
formula \cite{mank2,mank1,su}.

In the present paper we intend to study the connection of the quantum
deformation and quantum dissipation, by setting a master equation for the
deformed harmonic oscillator in the presence of a dissipative environment,
which is shown to be the deformed version of the master equation obtained in
the framework of the Lindblad theory for open quantum systems \cite{l1}.
When the deformation becomes zero, we recover the Lindblad master equation
for the damped harmonic oscillator \cite{ss,rev}. We are interested in
describing the role of nonlinearities which appear in the master equation, this
goal being motivated by the fact that the $q$-oscillator can be considered as a
physical system with a specific nonlinearity, called $q$-nonlinearity
\cite{mank2,mank1}. For a certain choice of the environment coefficients, a
master equation for the damped deformed oscillator has also been derived by
Mancini \cite{manc}. In Ref. \cite{eli}, Ellinas used the $q$-deformed oscillator
for treatments of dissipation of a two-level atom and of a laser mode.

The paper is organized as follows. In Sec. 2 we remind the basics
about the generalized deformed quantum oscillator, in particular
the $f$-oscillator and $q$-oscillator. Using a variant of the
Mancini's model \cite{manc}, in Sec. 3 we derive a master equation
for the $f$-deformed oscillator in the presence of a dissipative
environment. The equations of motion obtained for different
observables present a strong dependence on the deformation. Then
in Sec. 4 we write and solve in the stationary state the equation
for the density matrix in the number representation. In the
particular case when the environment is a thermal bath, we obtain
an expression for the equilibrium energy of the oscillator in the
approximation of a small deformation parameter. A summary and
conclusions are given in Sec. 5.

\section{Deformed quantum oscillators}

It is known that the ordinary operators $\{1, a, a^\dagger, N\}$
form the Lie algebra of the Heisenberg-Weyl group and the linear
harmonic oscillator can be connected with the generators of the
Heisenberg-Weyl Lie group. The generalized deformed quantum
oscillators \cite{das1,das2} are defined as the algebra generated
by the operators $\{1,A,A^\dagger,N\}$ and the structure function
$F(N),$  which satisfy the following relations: \bea
[A,N]=A,~~[A^\dagger,N]=-A^\dagger \label{com1}\eea and \bea
AA^\dagger =F(N+1),~~A^\dagger A=F(N),\label{rel}\eea where $F(N)$
is a positive analytic function with $F(0)=0$ and  $N$ is the
Hermitian number operator. It follows that the following
commutation and anticommutation relations are satisfied: \bea
[A,A^\dagger]=F(N+1)-F(N),~~
\{A,A^\dagger\}=F(N+1)+F(N).\label{com3}\eea The structure
function is a characteristics of the deformation. The number
operator $N$ is not equal to $A^\dagger A$ as in the ordinary
case. For $F(N)=N$ one obtains the relations for the usual
harmonic oscillator. Another choice is \bea F(N)={q^N-q^{-N}\over
q-q^{-1}}\equiv [N],\eea where the dimensionless $c$-number $q$ is
the deformation parameter. Then the operators $A$ and $A^\dagger$
are called $q$-deformed boson annihilation and creation operators
\cite{bied,macf}. For the $q$-deformed harmonic oscillator the
relations (\ref{rel}) become \bea A A^\dagger=[N+1],~~A^\dagger
A=[N],\eea with the commutation relation \bea
[A,A^\dagger]=[N+1]-[N].\label{com2}\eea If $q$ is real and
positive, then \bea [N]={\sinh(N\ln q)\over \sinh(\ln q)}\eea and
the condition of Hermitian conjugation $(A^\dagger)^\dagger=A$ is
satisfied.  In addition to the commutation relation, there exists
for the $q$-deformed oscillator the reordering relation \bea
AA^\dagger-q^{\mp 1}A^\dagger A=q^{\pm N},\label{com}\eea which is
usually taken as the definition of $q$-oscillators.

In the limit $q\to 1,$ $q$-operators tend to the ordinary
operators because $\lim_{q\to 1} [N]=N.$ Then Eqs. (\ref{com2})
and (\ref{com}) go to the usual boson commutation relation
$[A,A^\dagger]=1.$

The $q$-deformed boson operators $A$ and $A^\dagger$ can be
expressed in terms of the usual boson operators $a$ and
$a^\dagger$ (satisfying $[a,a^\dagger]=1, $ $N=a^\dagger a$ and
$[a,N]=a,~~[a^\dagger,N]=-a^\dagger$) through the relations
\cite{kuli,song}: \bea A=\sqrt{[N+1]\over N+1}a=a\sqrt{[N]\over
N},~~ A^\dagger=a^\dagger  \sqrt{[N+1]\over N+1}=\sqrt{[N]\over
N}a^\dagger .\eea

Using a nonlinear map \cite{poly,curt}, the $q$-oscillator has been interpreted
\cite{mank2,mank1} as a nonlinear oscillator with a special type of nonlinearity
which classically corresponds to an energy dependence of the oscillator
frequency. Other nonlinearities can also be introduced by making the frequency
to depend on other constants of motion, different from energy, through a
deformation function $f$ \cite{mank2,mank3}. Let us define the $f$-deformed
oscillator operators \cite{mank3} \bea
A=af(N)=f(N+1)a,~~A^\dagger=f(N)a^\dagger=a^\dagger
f(N+1),\label{def}\eea where $N=a^\dagger a.$ They satisfy relations
(\ref{com1}) and the commutation relation \bea
[A,A^\dagger]=(N+1)f^2(N+1)-Nf^2(N).\eea The function $f$ has a
dependence on the deformation parameter such that when the deformation
disappears, then $f\to 1$ and the usual algebra is recovered. Without loss of
generality, $f$ can be chosen real and nonnegative and it is reasonable from
the physical point of view to assume \cite {ani} that $f(0)=1$ and $f(N)=1$ for
a suitable large $N$. A deformation function depending on Laguerre polynomials
has been used in \cite{mat}. A different type of deformation has been
considered by Sudarshan \cite{sud}, who introduced the so-called harmonious
states. Other examples of deformed oscillators are connected to the excited
coherent states introduced by Agarwal \cite{aga} and Dodonov \cite{dod}. The
transformation (\ref{def}) from the operators $a,a^\dagger$ to
$A,A^\dagger$ represents a nonlinear non-canonical transformation, since it
does not preserve the commutation relation. The notion of $f$-oscillators
generalizes the notion of $q$-oscillators. Indeed, if \bea f(N)=\sqrt{[N]\over
N}=\sqrt{\sinh(N\ln q)\over N\sinh(\ln q)}, \label{qdef}\eea then the
operators $A,A^\dagger$ in Eqs. (\ref{def}) satisfy the $q$-deformed
commutation relations (\ref{com2}). This means that a Hamiltonian operator of
the form $A^\dagger A=f(N)a^\dagger a f(N)$ has a spectrum with the same
structure as the spectrum of $a^\dagger a.$ The difference is that the
eigenvalues in the basis of the Fock space are $nf^2(n),$ $n=0,1,2,...,$
instead of $n.$ This spectrum associated with $q$-deformation grows with $n$
like $\sinh(n\ln q),$ i. e. exponentially for large occupation numbers $n,$
compared to the ordinary case, in which the spectrum is equidistant. The
Hamiltonian of the $f$-deformed harmonic oscillator is ($\omega$ is the
ordinary frequency) \bea {\cal H}={\hbar\omega\over
2}(AA^\dagger+A^\dagger A) ={\hbar\omega\over
2}[(N+1)f^2(N+1)+Nf^2(N)] \label{ham}.\eea It is diagonal on the
eigenstates  $|n>$ and in the Fock space its eigenvalues are \bea
E_n={\hbar\omega\over 2}[(n+1)f^2(n+1)+nf^2(n)].\label{eig}\eea In the
limit $f\to 1$ ($q\to 1$ for $q$-oscillators), we recover the ordinary expression
$E_n=\hbar\omega(n+{1/2}).$

Using the operator Heisenberg equation with the Hamiltonian $\cal $
(\ref{ham}) \bea i\hbar{da\over dt}=[{a,\cal H}]\eea or the evolution
operator $U(t)=\exp[-(i/\hbar){\cal H}(N)t],$ we obtain the following solutions
to the Heisenberg equations of motion for the operators $a$ and $a^\dagger $
\cite{manc,mank3}: \bea
a(t)=\exp[-i\omega\Omega(N)t]a,~~a^\dagger(t)=a^\dagger \exp[i\omega
\Omega(N)t],\label{temp}\eea where \bea \Omega(N)={1\over
2}[(N+2)f^2(N+2)-Nf^2(N)].\eea For a $q$-deformed harmonic oscillator,
\bea \Omega(N)={1\over 2}([N+2]-[N])\eea and for a small deformation
parameter $\tau$ ($\tau=\ln q$), \bea \Omega(N)=1+{\tau^2\over
2}(N+1)^2.\label{deform}\eea

\section{Quantum Markovian master equation}

In order to discuss the dynamics of the open system S, we use a microscopic
description of the composite system S+B. As the subsystem S of interest we
take the $f$-deformed harmonic oscillator with the Hamiltonian $\cal H$
(\ref{ham}), and B is the environment (bath) with the Hamiltonian $H_B.$ The
coupled system with the total Hamiltonian $H_T={\cal H}+H_B+V$ ($V$ is the
interaction Hamiltonian) is described by a density operator $\chi(t),$ which
evolves in time according to the von Neumann-Liouville equation \bea
{d\chi(t)\over dt}=-{i\over \hbar}[H_T,\chi(t)].\eea When the Hamiltonian
evolution of the total system is projected onto the space of the harmonic
oscillator, the reduced density operator of the subsystem is given by \bea
\rho(t)={\rm Tr_B}\chi(t).\label{def1}\eea The derivation of the reduced
density operator in which the operators of the environment system have been
eliminated up to second order of the perturbation theory can be taken from
literature \cite{loui,hake,gard,carm,wal}. We assume that the interaction
potential $V$ is linear in the coordinate operator $s_1=q$ and momentum
operator $s_2=p$ in the Hilbert space of the subsystem. Then, following
\cite{carm}, we can write down the master equation for the density operator of
the open quantum system in the Born-Markov approximation:
\bea {d\rho(t)\over dt}=-{i\over \hbar}[{\cal H},\rho(t)]
+{1\over\hbar^2}\sum_{i,j=1,2}\int_0^t dt'\{
C_{ij}^*(t')[s_i,\rho(t)s_j(-t')]
+C_{ij}(t')[s_j(-t')\rho(t),s_i]\},\label{mast}\eea where the coefficients
$C_{ij}(t)$ are correlation functions of the environment operators. It is
assumed that the correlation functions decay very rapidly on the time scale on
which $\rho(t)$ varies. Ideally, we might take $C_{ij}(t')\sim \delta(t').$ The
Markov approximation relies on the existence of two widely separated time
scales: a slow time scale for the dynamics of the system S and a fast time scale
characterizing the decay of environment correlation functions \cite{carm}.

In order to get the time dependence of the operators $s_1(t)=q(t)$ and
$s_2(t)=p(t)$, we express them through the relations ($m$ is the oscillator
mass) \bea q(t)=\sqrt{\hbar\over 2m\omega}(a^\dagger(t)+a(t)), ~~
p(t)=i\sqrt{\hbar m\omega\over 2}(a^\dagger(t)-a(t)) \eea and then insert
Eq. (\ref{temp}) for $a(t)$ and $a^\dagger(t).$  Then the master equation
results \bea {d\rho(t)\over dt}=
-{i\over \hbar}[{\cal H},\rho(t)]\nonumber\\
+{1\over 2\hbar^2}\int_0^t dt'
\{C_{11}^*(t')[q,\rho(t)(qE_-+E_+q-{i\over m\omega}(pE_--E_+p))]\nonumber\\
+C_{11}(t')[(qE_-+E_+q-{i\over m\omega}(pE_--E_+p))\rho(t),q]\nonumber\\
+iC_{22}^*(t')[p,\rho(t)(m\omega(qE_--E_+q)-i(pE_-+E_+p))]\nonumber\\
+iC_{22}(t')[(m\omega(qE_--E_+q)-i(pE_-+E_+p))\rho(t),p]\nonumber\\
+iC_{12}^*(t')[q,\rho(t)(m\omega(qE_--E_+q)-i(pE_-+E_+p))]\nonumber\\
+C_{21}(t')[(qE_-+E_+q-{i\over m\omega}(pE_--E_+p))\rho(t),p]\nonumber\\
+C_{21}^*(t')[p,\rho(t)(qE_-+E_+q-{i\over m\omega}(pE_--E_+p))]\nonumber\\
+iC_{12}(t')[(m\omega(qE_--E_+q)-i(pE_-+E_+p))\rho(t),q]\},
\label{mast1}\eea
where  we have introduced the following notations:
\bea E_+=\exp[i\omega\Omega(N)t'],~~E_-=\exp[-i\omega\Omega(N)t'].\eea

If the environment is sufficiently large we may assume that the time correlation
functions decay fast enough to zero for times $t'$ longer than the relaxation
time $t_B$ of the environment: $t'\gg t_B.$ Therefore, if we are interested in
the dynamics of the subsystem over times which are longer than the
environment relaxation time,  $t\gg t_B,$ we may use the Markov
approximation and replace the upper limit of integration $t$ by $\infty.$
Physically, this amounts to assuming that the memory functions
$C_{ij}(t')E_{\pm}(t')$ decay over a time which is much shorter than the
characteristic evolution time of the system of interest. After certain assumptions
\cite{loui,wal}, one can define the complex decay rates, which govern the rate
of relaxation of the system density operator as follows: \bea \int_0^\infty dt'
C_{11}(t')E_{+}= \int_0^\infty dt'
C_{11}^*(t')E_{+}=D_{pp}(\Omega),\label{dif1}\eea \bea \int_0^\infty
dt'C_{22}(t')E_{+}= \int_0^\infty
dt'C_{22}^*(t')E_{+}=D_{qq}(\Omega),\eea \bea \int_0^\infty
dt'C_{12}(t')E_{+}=\int_0^\infty dt'C_{21}^*(t')E_{+}
=-D_{pq}(\Omega)+{i\hbar\over 2}\lambda(\Omega),\label{disco}\eea with
$D_{pp}(\Omega)>0,$ $D_{qq}(\Omega)>0$ and \bea
D_{pp}(\Omega)D_{qq}(\Omega)-D_{pq}^2(\Omega)\ge {\hbar^2\over
4}{\lambda}^2(\Omega).\label{posi}\eea In fact, $D_{pp}(\Omega),$
$D_{qq}(\Omega),$ $D_{pq}(\Omega)$ and $\lambda(\Omega)$ play the
role of deformed diffusion and, respectively,  dissipation coefficients and the
relation (\ref{posi}) ensures the positivity of the density operator. The
existence of these coefficients reflects the fact that, due to the interaction, the
energy of the system is dissipated into the environment, but noise arises also
(in particular, thermal noise), since the environment also distributes some of its
energy back to the system. In addition, we assume in the following
$\lambda(\Omega)=\lambda=const.$ Then the master equation (\ref{mast1})
for the damped deformed harmonic oscillator takes the form
\bea   {d    \rho \over dt}=-{i\over \hbar}[{\cal H},    \rho]\nonumber\\
+{1\over 2\hbar^2}\{[(\{D_{pp}(\Omega),q\}+{i\over m\omega}[D_{pp}(\Omega),p])
\rho,q]+[(\{D_{qq}(\Omega),p\}-im\omega[D_{qq}(\Omega),q])\rho,p]\nonumber\\
+[(m\omega[iD_{pq}(\Omega)+{\hbar\over 2}\lambda,q]
-\{D_{pq}(\Omega)-{i\hbar\over 2}\lambda,p\})\rho,q]\nonumber\\
-[({1\over m\omega}[iD_{pq}(\Omega)-{\hbar\over 2}\lambda,p]
+\{D_{pq}(\Omega)+{i\hbar\over 2}\lambda,q\})\rho,p]
+H.c.\}.\label{mast2}\eea We notice that the deformation is present in both
the commutator containing the oscillator Hamiltonian $\cal H,$ as well as in the
dissipative part of the master equation, which describes the influence of the
environment on the deformed oscillator. This master equation preserves the
Hermiticity property of the density operator and the normalization (unit trace)
at all times, if at the initial time it has these properties. In the limit $f\to 1$
$(\Omega\to 1),$ the deformation disappears and Eq. (\ref{mast2}) becomes
the Markovian master equation for the damped harmonic oscillator, obtained in
the Lindblad theory for open quantum systems, based on completely positive
dynamical semigroups \cite{l1,ss,rev}.

Expressing the coordinate and momentum operators back in terms of the
creation and annihilation operators and introducing the notations \bea
D_+(\Omega)\equiv {1\over 2\hbar}[m\omega D_{qq}(\Omega)+
{D_{pp}(\Omega)\over m\omega}],~~D_-(\Omega)\equiv {1\over
2\hbar}[m\omega D_{qq}(\Omega)-{D_{pp}(\Omega)\over m\omega}],\eea
the master equation (\ref{mast2}) for the damped deformed harmonic
oscillator takes the form: \bea   {d    \rho \over dt}=-{i\over \hbar}[{\cal
H},    \rho] \nonumber \\
+\{[[D_+(\Omega)a,\rho],a^\dagger]-[[a^\dagger(D_-(\Omega)
+{i\over\hbar}D_{pq}(\Omega)),\rho],a^\dagger]-{\lambda\over 2}
[a^\dagger,\{a,\rho\}]+H.c.\}.\label{mast5}\eea Mancini considered
in Ref. \cite{manc} a squeezed bath for the dynamics of the damped
deformed harmonic oscillator and his model can be recovered by
taking the following coefficients in Eq. (\ref{mast5}): \bea
D_+(\Omega)=\gamma(N+{1\over 2}),
~~D_-(\Omega)+{i\over\hbar}D_{pq}(\Omega)=-\gamma M,
~~\lambda=\gamma. \eea

In the particular case of a thermal equilibrium of the bath at temperature $T$
($k$ is the Boltzmann constant), we take the diffusion coefficients of the form
(in concordance with Mancini's results \cite{manc}) \bea m\omega
D_{qq}(\Omega)={D_{pp}(\Omega)\over m\omega}= {\hbar \over
2}\lambda\coth{\hbar\omega\Omega\over 2kT},~~D_{pq}(\Omega)=0.
\label{defth}\eea In the limit $\Omega\to 1,$ the deformed diffusion
coefficients (\ref{defth}) take the known form obtained for the damped
harmonic oscillator in the particular case when the asymptotic state is a Gibbs
state \cite{ss,rev}: \bea D_{pp}={\hbar m\omega\over
2}\lambda\coth{\hbar\omega\over 2kT}, ~~D_{qq}={\hbar\over
2m\omega}\lambda\coth{\hbar\omega\over 2kT},
~~D_{pq}=0.\label{coegib} \eea

The meaning of the master equation becomes clear when we transform it into
equations satisfied by the expectation values of observables involved in the
master equation, $<O>={\rm Tr}[\rho(t)O],$ where $O$ is the operator
corresponding to such an observable. We give an example, multiplying both
sides of Eq. (\ref{mast5}) by the number operator $N$ and taking the trace. In
the case of a thermal bath, with the diffusion coefficients (\ref{defth}), the
equation of motion for the expectation value of $N$ has the form \bea{d\over
dt}<N>=\lambda[<(\coth{\hbar\omega\Omega(N)\over 2kT}-1)(N+1)>
-<(\coth{\hbar\omega\Omega(N-1)\over 2kT}+1)N>].\label{expv}\eea This
equation leads to a time dependence of the number of quanta on dissipation
and temperature, compared to the case of an oscillator without dissipation,
where the number of quanta is conserved. We remark that in the case of a
thermal bath at $T=0,$ Eq. (\ref{expv}) takes the form \bea{d\over
dt}<N>=-2\lambda<N>,\eea so that the average number of quanta $<N>$
does not depend on deformation, it only decreases exponentially with
dissipation.

We consider  another example, taking the simplest case of a
thermal bath at $T=0,$ when both diffusion and dissipation
coefficients do not depend on the deformation,
$D_+=\lambda/2=const.$ Even in this situation, the equations of
motion for the expectation values are yet complicated, because
they do not form a closed system. Multiplying both sides of Eq.
(\ref{mast5}) by the operators $a$ and, respectively, $\Omega(N)a$
and taking throughout the trace, we get the following equations
for the expectation values of these operators: \bea{d\over
dt}<a>=-i\omega<\Omega(N)a>-\lambda<a
>,\eea \bea{d\over
dt}<\Omega(N)a>=-i\omega<\Omega^2(N)a>+\lambda<[2N\Omega(N-1)-
(2N+1)\Omega(N)]a >.\eea These examples show that the equations of motion
contain nonlinearities introduced by the deformed Hamiltonian $\cal H$ and,
therefore, depend on the deformation function.

\section{Equation for the density matrix of the damped deformed oscillator}

Let us rewrite the master equation (\ref{mast5}) for the density matrix by
means of the number representation. Specifically, we take the matrix elements
of each term between different number states denoted by $|n>$, and using
$N|n>=n|n>,$ $a^+|n>=\sqrt{n+1}|n+1>$ and $a| n>=\sqrt n| n-1>,$ we get
\newpage
\bea {d\rho_{mn}\over dt}=-{i\omega\over 2}[mf^2(m)+(m+1)f^2(m+1)
-nf^2(n)-(n+1)f^2(n+1)]\rho_{mn}\nonumber\\
-[(m+1)D_+(\Omega(m))+mD_+(\Omega(m-1))
+(n+1)D_+(\Omega(n))+nD_+(\Omega(n-1))
-\lambda]\rho_{mn}\nonumber\\
+\sqrt{(m+1)(n+1)}[D_+(\Omega(m))+D_+(\Omega(n))+\lambda]
\rho_{m+1,n+1}\nonumber\\
+\sqrt{mn}[D_+(\Omega(m-1))+D_+(\Omega(n-1))-\lambda]\rho_{m-1,n-1}\nonumber\\
-\sqrt{(m+1)n}[D_-(\Omega(m))+D_-(\Omega(n-1))-{i\over\hbar}(D_{pq}(\Omega(m))
+D_{pq}(\Omega(n-1)))]\rho_{m+1,n-1}\nonumber\\
-\sqrt{m(n+1)}[D_-(\Omega(m-1))+D_-(\Omega(n))+{i\over\hbar}(D_{pq}
(\Omega(m-1))
+D_{pq}(\Omega(n)))]\rho_{m-1,n+1}\nonumber\\
+\sqrt{(m+1)(m+2)}[D_-(\Omega(m+1))-{i\over \hbar}D_{pq}(\Omega(m+1))]
\rho_{m+2,n}
\nonumber\\
+\sqrt{(n+1)(n+2)}[D_-(\Omega(n+1))+{i\over \hbar}D_{pq}(\Omega(n+1))]
\rho_{m,n+2}\nonumber\\
+\sqrt{m(m-1)}[D_-(\Omega(m-2))+{i\over \hbar}D_{pq}(\Omega(m-2))]\rho_{m-2,n}
\nonumber\\
+\sqrt{n(n-1)}[D_-(\Omega(n-2))-{i\over
\hbar}D_{pq}(\Omega(n-2))]\rho_{m,n-2}. \label{numeq}\eea Here, we
have used the abbreviated notation $\rho_{mn}=<m\vert\rho(t)\vert n>.$
This equation gives an infinite hierarchy of coupled equations for the matrix
elements. When \bea D_-(\Omega(n))=0,~~D_{pq}(\Omega(n))=0,\eea the
diagonal elements are coupled only amongst themselves and not coupled to the
off-diagonal elements. In this case the diagonal elements (populations) satisfy a
simpler set of master equations: \bea {dP(n)\over
dt}=-[2(n+1)D_+(\Omega(n))+2nD_+(\Omega(n-1))
-\lambda]P(n)\nonumber\\
+(n+1)[2D_+(\Omega(n))+\lambda]P(n+1)
+n[2D_+(\Omega(n-1))-\lambda]P(n-1),\label{pop}\eea where we have
set $P(n)\equiv\rho_{nn}.$ We define the transition probabilities
\bea t_+(n)=(n+1)[2D_+(\Omega(n))-\lambda],
~~t_-(n)=n[2D_+(\Omega(n-1))+\lambda].\eea With these notations
Eq. (\ref{pop}) becomes: \bea {dP(n)\over
dt}=t_+(n-1)P(n-1)+t_-(n+1)P(n+1)-[t_+(n)+t_-(n)]P(n)
\label{pop1}.\eea The steady state solution of Eq. (\ref{pop1}) is
found to be \bea P_{ss}(n)=P(0)\prod_{k=1}^{n}{2D_+(\Omega(k-1))-
\lambda\over 2D_+(\Omega(k-1))+\lambda}.\eea We note that in the
steady state the detailed balance condition holds: \bea
t_-(n)P(n)=t_+(n-1)P(n-1).\eea

In the particular case of a thermal state, when the diffusion coefficients have
the form (\ref{defth}), the stationary solution of Eq. (\ref{pop1}) takes the
following form: \bea P_{ss}^{th}(n)=Z_f^{-1}\exp\{-{\hbar\omega\over 2
kT} [(n+1)f^2(n+1)+nf^2(n)]\}\label{popth},\eea where \bea
Z_f^{-1}=P(0)\exp{\hbar\omega f^2(1)\over 2kT}\eea and $Z_f$ is the
partition function: \bea Z_f=\sum_{n=0}^{\infty}\exp\{-{\hbar\omega\over
2 kT} [(n+1)f^2(n+1)+nf^2(n)]\}\label{part}.\eea In his model, Mancini
obtained a result similar with Eq. (\ref{popth}).

Using Eq. (\ref{eig}), the distribution (\ref{popth}) can be written \bea
P_{ss}^{th}(n)=Z_f^{-1}\exp(-{E_n\over kT}).\eea Expression
(\ref{popth}) represents the Boltzmann distribution for the deformed harmonic
oscillator. In the limit $f\to 1$ the probability $P_{ss}^{th}(n)$ becomes the
Boltzmann distribution for the ordinary harmonic oscillator with the well-known
partition function \bea Z={1\over 2\sinh{\hbar\omega\over 2kT}}.\eea

For the $q$-oscillator described by the Hamiltonian $\cal H$ (\ref{ham}) and
weakly coupled to a reservoir kept at the temperature $T,$ the $q$-deformed
partition function can be obtained as a particular case of the partition function
$Z_f$ (\ref{part}), by taking the deformation function (\ref{qdef}): \bea
Z_q=\sum_{n=0}^\infty\exp\{-{\hbar\omega\over 2kT}{\sinh(\tau(n+1))+
\sinh(\tau n)\over \sinh \tau}\}.\eea In the limit of a small deformation $\tau$
we can write \cite{mank1} $Z_q=Z+b\tau^2,$ where \bea b=-{\beta Z\over
12}(2\bar{n^3}+3\bar{n^2}+\bar n),\eea with \bea \bar n={1\over
e^\beta-1},~~ \bar{n^2}={e^\beta+1\over (e^\beta-1)^2},~~
\bar{n^3}={e^{2\beta}+4e^\beta+1\over (e^\beta-1)^3},~~
\beta={\hbar\omega\over kT}.\eea We can calculate the equilibrium energy by
using the formula \bea E(\infty)=-\hbar\omega{1\over Z_q}{\partial Z_q\over
\partial\beta}\eea and obtain \bea E(\infty)={\hbar\omega\over
2}(\coth{\hbar\omega\over 2kT}+{\tau^2}c),\eea where \bea
c={e^{\beta}\over (e^\beta-1)^2}[{e^\beta+1\over e^\beta-1}
-\beta{e^{2\beta}+4e^\beta+1\over (e^\beta-1)^2}].\eea We note that in
the approximation of a small deformation parameter $\tau,$ the energy of the
deformed damped oscillator depends on the oscillator ground state energy
$\hbar\omega/2$ and on the temperature $T.$ Evidently, when there is no
deformation $(\tau\to 0),$ one recovers the energy of the ordinary harmonic
oscillator in a thermal bath \cite{ss,rev}. In the limit $T\to 0,$ one has $c\to
0,$ $E(\infty)=\hbar\omega/2$ and the deformation does not play any role.

\section{Summary and conclusions}

Our purpose was to study the dynamics of the deformed quantum
harmonic oscillator in interaction with a dissipative environment,
in particular with a thermal bath. We derived in the Born-Markov
approximation a master equation for the reduced density operator
of the damped $f$-deformed oscillator. The one-dimensional $f$- or
$q$-oscillator is a nonlinear quantum oscillator with a specific
nonlinearity and, consequently, the diffusion and dissipation
coefficients which model the influence of the environment on the
deformed oscillator depend strongly on the introduced
nonlinearities. The equations of motion for the observables of the
considered system are also nonlinear. In the limit of zero
deformation, the master equation for the deformed damped
oscillator takes the form of a master equation for the damped
oscillator obtained in the framework of the Lindblad theory of
open quantum systems based on quantum dynamical semigroups. We
have also derived the equation for the density matrix in the
number representation. In the case of a thermal bath we obtained
the stationary solution, which is the Boltzmann distribution for
the deformed harmonic oscillator. In the approximation of a small
deformation, we obtained the expression of the equilibrium energy
of the deformed harmonic oscillator, which depends on the
oscillator ground state energy and on the temperature.

The master equation for the damped deformed harmonic oscillator is
an operator equation. It could be useful to study its consequences
for the density operator by transforming this equation into more
familiar forms, such as partial differential equations of
Fokker-Planck type for the Glauber, antinormal ordering and Wigner
quasiprobability distributions or for analogous deformed
quasiprobabilities \cite{ani} associated with the density
operator. It could also be interesting to find the states which
minimize the rate of entropy production for the damped deformed
harmonic oscillator. In the case of the undeformed damped
oscillator such states are represented by correlated coherent
states. For the damped deformed oscillator the corresponding
states could be deformed (nonlinear) coherent states \cite{mank3},
which may play an important role in the description of the
phenomenon of environment induced decoherence. The dissipative
dynamics of deformed coherent states superposition and the related
coherence properties have been studied recently by Mancini and
Man'ko \cite{manc1}.

{\bf Acknowledgements}

One of us (A. I.) is pleased to express his sincere gratitude for the hospitality at
the Institut f\"ur Theoretische Physik in Giessen and support by BMFT. A. I. also
gratefully acknowledges the support from the Romanian Academy under grant
50/2000.

\end{document}